\documentclass[journal=jctcce,manuscript=article]{achemso}
\usepackage[numbers,sort&compress]{natbib}
\usepackage{textcomp}
\usepackage{physics}
\usepackage{gensymb}
\usepackage{graphicx}
\usepackage{lineno}
\usepackage{amsmath}
\usepackage{upgreek}
\usepackage{multirow}
\usepackage{array}
\usepackage{booktabs}
\usepackage{multicol}
\usepackage{booktabs}
\usepackage{cleveref}
\graphicspath{{./figures/}}

\author{Jianlan Ye}
\affiliation[Arizona State University]
{School for the Engineering of Matter, Transport and Energy, Arizona State
University, Tempe, AZ, 85287, United States}
\author{Vipin Agrawal}
\author{Yiyang Li}
\affiliation[Arizona State University]
{School for the Engineering of Matter, Transport and Energy, Arizona State
University, Tempe, AZ, 85287, United States}
\author{Minghao Liu}
\affiliation[Arizona State University]
{School for the Engineering of Matter, Transport and Energy, Arizona State
University, Tempe, AZ, 85287, United States}
\author{Jing Hu}
\affiliation[Arizona State University]
{School for the Engineering of Matter, Transport and Energy, Arizona State
University, Tempe, AZ, 85287, United States}
\author{Jay Oswald}
\email{joswald1@asu.edu}
\affiliation[Arizona State University]
{School for the Engineering of Matter, Transport and Energy, Arizona State 
University, Tempe, AZ, 85287, United States}

\title{KDE Based Coarse--graining of Semicrystalline Systems with Correlated Three--body Intramolecular Interaction}
\begin{document}
\begin{abstract}
\noindent
We present an extension to the iterative Boltzmann inversion method to generate coarse--grained models with three--body intramolecular potentials that can reproduce correlations in structural distribution functions.
The coarse--grained structural distribution functions are computed using kernel density estimates to produce analytically differentiable distribution functions with controllable smoothening via the kernel bandwidth parameters.
Bicubic interpolation is used to accurately interpolate the three--body potentials trained by the method.
To demonstrate this new approach, a coarse--grained model of polyethylene is constructed in which each bead represents an ethylene monomer.
The resulting model reproduces the radial density function as well as the joint probability distribution of bond--length and bond--angles sampled from target atomistic simulations with only a 10\% increase in the computational cost compared to models with independent bond--length and bond--angle potentials.
Analysis of the predicted crystallization kinetics of the model developed by the new approach reveals that the bandwidth parameters can be tuned to accelerate the modeling of polymer crystallization.
Specifically, computing target RDF with larger bandwidth slows down the secondary crystallization and increasing the bandwidth in $\theta$--direction of bond--length and bond--angle distribution reduces the primary crystallization rate.
\end{abstract}

\section{Introduction}

Coarse--grained (CG) molecular dynamics (MD) simulations are commonly used to study polymer physics for their ability in probing molecular processes at substantially longer time and length scales than all--atom models.\cite{nielsen2004coarse,peter2009multiscale,liu2019coarse,liu2023coarse}
While including more atoms in a bead reduces the computational cost of the model, the reduced resolution weakens its ability to accurately represent the structural properties of the underlying polymer.
For example, Salero~et~al.\cite{salerno2016resolving} found that polyethylene (PE) models with flexible chains did not crystallize at all when the coarse--grained beads represent more than four methylene groups. 
They attributed the phenomenon to the exceedingly large equilibrium bond--length compared with the equilibrium pair distance.\cite{hoy2013simple,nguyen2015effect}
At the lowest degree of coarse--graining, united--atom model (UAM) that lumps hydrogen atoms with their bonded carbon atoms along the chain backbone are more capable at reproducing structural features.
The UAM of polyethylene, initially developed by Paul et al.\cite{paul1995optimized} and subsequently modified by Waheed et al.,\cite{waheed2002molecular, waheed2005molecular} yields a hexagonal crystal structure.\cite{yi2013molecular}
The UAM developed by Zubova et al. reproduced the expected orthorhombic crystal structure by adjusting the displacements of beads from the chain axes and the van der Waals radii.\cite{zubova2017coarse,strelnikov2017coarse}
However, given that united atom models have a relatively small reduction in the degrees of freedom compared to all--atom models, the range of time--scales the models can access remains highly limited.
Therefore, many MD studies of polyethylene crystallization\cite{waheed2002molecular, Lavine2003, Yamamoto2004, Ko2004, Yamamoto2013, paajanen2019crystallization, sliozberg2018ordering} that employed UAMs used systems with chain lengths no longer than 1000 methyl units because the drastically reduced crystallization rate with longer chains necessitates unfeasible simulation time, although real polyethylene chains are much longer.
Thus, developing CG models with efficient computation and accurate representation of structure can strengthen the capability of MD simulations in the investigations that require large systems and long time scales.

It has been shown that the accuracy of CG models in representing water,\cite{larini2010multiscale} RNA,\cite{poursina2011strategies} and proteins\cite{munson1997statistical,li2005geometric} increase when considering multi--body interactions.
Larini et al.\cite{larini2010multiscale} introduced three--body non--bonded interactions into their single--site water CG model by adding the relative angular orientation of the triplets of water molecules into the energy formulation; the addition greatly improved the model's ability to reproduce the radial distribution function (RDF) and angular distribution function (ADF).
Ejtehadi et al.\cite{ejtehadi2004three} and Krishna et al.\cite{krishna2010role} showed that three--body interactions play an important role in accurately representing the protein folding mechanism and the structure of HIV-1 capsid, respectively.
Also, Schaposchnikov et al.\cite{schapotschnikow2009understanding} showed that the contribution of three--body effect is 20\% to 40\% of ligand--capped nanocrystals.
Multibody interactions can also improve the representability of polymer structures. 
Fukunaga et al.\cite{fukunaga2002coarse} presented a CG model that groups three CH$_2$ units into a CG bead and showed that the bond--lengths and torsion angles distributions strongly correlated with the bond--angle distributions at room temperature.

CG model interactions, e.g. non--bonded interactions, can be described either by a functional form or interpolated from tabulated values.
Although, the tabulated potentials are more flexible in describing complex structures, they are also more prone to sampling noise.
The noise--induced energetic fluctuations causes magnified oscillations when computing the forces, leading to a reduced stable time step.
When training CG models with a structure--based method such as iterative Boltzmann inversion (IBI),\cite{reith2003deriving} the structural distribution functions, e.g. RDF, are commonly constructed using histograms, which has well--known disadvantages such as limited smoothness with a small sample size and sensitivity to the bin size.
Even though smoothing methods such as B-spline\cite{fukunaga2002coarse} and cubic spline\cite{larini2010multiscale} can be used to smooth the potential surface parametrized using histograms, McCabe et al.\cite{mccabe2014kernel} showed that with the kernel density estimation (KDE), we can readily obtain analytically differentiable and smooth structural distribution functions regardless of sample size.

In this paper, we develop an approach for constructing CG models with three--body interactions to reproduce correlated bond--length and bond--angle distribution functions.
The resulting potentials are generated on a two--dimensional grid and evaluated using bicubic interpolation.
To generate accurate and smooth energy derivatives, the KDE is used to calculate the structural distribution functions that enables analytical differentiation of energy.
We demonstrate the new method by developing coarse--grained models for PE, a model semicrystalline polymer.
The models are named as CG--BA model.
We also study how the KDE bandwidths affect the activation energy of diffusion, computational performance, and the crystallization kinetics of the resulting CG models.

\section{Methodology}

\subsection{CG model}
\begin{figure*}[tbp]
  \centering
  \includegraphics{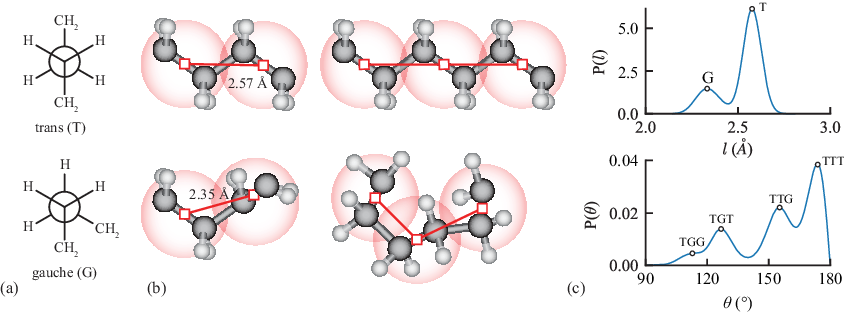}
  \caption{Schematic demonstrating how (a) underlying trans and gauche torsional conformations at the atomistic representation lead to (b) multiple bond--lengths and bond--angles at the coarse--grained representation, which (c) manifests as multi--peaked coarse--grained bond and angle distributions.}
  \label{fig:bdf}
\end{figure*}

Similar to the CG models developed by Li et al.,~\cite{li2019systematic} we use a coarse--grained mapping scheme in which each CG interaction site, hereafter referred to as a bead, represents an ethylene unit and each bead center is defined as the midpoint of the C--C bond.
Beads are considered bonded if they represent adjacent monomers on a molecular chain.
Due to the relatively fine scale of the mapping scheme and the predominant trans and gauche torsion conformations in PE, the coarse--grained bond--length distribution (BDF) is double--peaked, as shown in Figure~\ref{fig:bdf}, leading to correlations between the BDF and ADF of the CG model.

The CG models were developed using the IBI method to produce tabulated CG potentials. 
The total potential energy of the CG system includes the summation of a non--bonded potential and a bond--angle energy potential, written as:

\begin{equation}
    U_{total} = U_{nb} + U_{ba}.
\end{equation}
The non--bonded energy function is described by tabular values with a cutoff distance $r_c = 16$~\AA, and applies to all pairs of atoms (pair potential), except contributions from the first--,  second--, and third--bonded neighbors.
The bond--angle energy term (BA potential) includes three--body energy contributions from all sequential bead triplets along each chain and is also described by tabular values, written as:

\begin{equation}
    U_{ba} = \sum_{ijk}^{N_{\theta}} s_i \mathcal V(l_{ij}, \theta_{ijk}) 
                                   + s_k \mathcal V(l_{jk}, \theta_{ijk})
    \label{eq:BA_potential}
\end{equation}
where $s_i = 1$ if bead $i$ is an end bead and $s_i = \frac{1}{2}$ if bead $i$ is an internal bead within the chain. 
The energy contribution from internal beads is halved because the bonds in a linear chain are shared by adjacent angles.
It should be emphasized that for the CG model to be able to reproduce correlations between the bond--length and bond--angle distributions, the contributions of the bond--length and bond--angle energies cannot be additively decomposed, necessitating a multivariate potential energy function.
The bond--angle potential as defined in Eq.~(\ref{eq:BA_potential}) is implemented as a user angle style within the LAMMPS molecular dynamics code and is available along with the tabulated potentials on GitHub.\cite{CGBA2022}

\subsection{KDE--based probability distribution}
Following the work by McCabe\cite{mccabe2014kernel}, we computed the KDE of the RDF and the BADF.
The RDF is computed on a one--dimensional grid using the following:

\begin{equation}
    g(r_k) =  
      \frac{\frac{1}{N} \sum_i^N \sum_{j \in \mathcal N_i} K_r(r_k-r_{ij}) \Delta r} 
           {\frac{4\pi}{3}  \bar\rho
             \left((r_k+\frac{1}{2}\Delta r)^3 - (r_k-\frac{1}{2}\Delta r)^3\right) },
    \label{eq:kde_rdf}
\end{equation}
where $N$ is the number of beads, $\Delta r$ is the grid spacing, $r_{ij}$ is the distance between beads $i$ and $j$, $\mathcal N_i$ is the set of beads within the cutoff distance of bead $i$, $r_k$ is the grid point that starts at $\frac{\Delta r}{2}$, and $\bar\rho$ is the overall number density of beads in the system.
A Gaussian kernel function $K_r$ distributes the contribution of each pair distance to the grid value using a bandwidth parameter $w_r$:

\begin{equation}
    K_r(r) = \frac{1}{w_r \sqrt{2\pi}} \, 
            \exp \left(-\frac{r^2}{2w_r^2} \right)
    \label{eq:rdf_kernel}
\end{equation}
The numerator of (\ref{eq:kde_rdf}) is similar to a histogram of the number of beads located within spherical shells emanating out from a central bead, except that the counting of each bead is smeared across multiple grid points by the kernel function, defined in Eq.~(\ref{eq:rdf_kernel}).
The denominator produces the standard normalization of the RDF.

The joint bond--length and bond--angle probability density distribution function (BADF) is computed using a multidimensional KDE. 
We introduced two measures for the probability function, $P(l, \theta)$ is the true probability density function while $\widehat{P}(l, \theta)$ is the entropy--scaled probability.
\begin{align}
    P(l, \theta) = \frac{1}{N} \sum \limits_{I=1}^{N}
               K_H (\mathbf{x}-{{\mathbf{x}}_{I1}}) + K_H (\mathbf{x}-{{\mathbf{x}}_{I2}})\\
    \widehat{P}(l, \theta) = \frac{1}{N} \sum \limits_{I=1}^{N}
               \frac{K_H (\mathbf{x}-{{\mathbf{x}}_{I1}}) + K_H (\mathbf{x}-{{\mathbf{x}}_{I2}})}{\sin \theta_I}
    \label{eq:badf}
\end{align}
where $\mathbf{x}_{I1} = (l_{I1}, \theta_I)$ and $\mathbf{x}_{I2} = (l_{I2}, \theta_I)$ are split from triplet $(l_{I1}, l_{I2}, \theta_I)$ sampled from the systems. 
This decomposition requires $l_{I1}$ and $l_{I2}$ to be uncorrelated. 
The Spearman correlation coefficient of the CG bond lengths is 0.027, indicating that the assumption is valid for the coarse--grained model considered here.
The kernel function used for computing the BADF, denoted by $K_H$, is a two--dimensional Gaussian function, defined as:
\begin{equation}
  K_H(\mathbf{x}) = \frac{1}{2\pi {\left| \mathbf{H} \right|}^{1/2}}
                    \exp \left(-\frac{\mathbf x^T \cdot \mathbf H^{-1} \cdot \mathbf x}{2} \right),
    \label{eq:badf_kernel}
\end{equation}
where $\mathbf{H}$ is a matrix,
\begin{equation}
  \mathbf{H} = \left[\begin{matrix}
     w_l^2 & 0  \\
     0     & w_\theta^2  
  \end{matrix} \right], 
\end{equation}
in which, $w_l$ and $w_{\theta}$ are the bond--length and bond--angle bandwidth parameters, respectively. 

Since the Jacobian of the transformation from Cartesian to polar coordinates is proportional to $\sin\theta$, the probability distribution $P(l,\theta)$ vanishes at $\theta = \pi$, which introduces numerical artifacts to the IBI steps that worsen with increasing $w_\theta$. 
An entropy correction factor $\sin \theta_I$ is introduced to the kernel function to account for this entropic effect as shown in Eq.~\eqref{eq:badf}.
The computed values of the entropy--scaled BADF spuriously decrease as $\theta$ approaches $\pi$.
To mitigate this issue, mirrored data points, i.e., $(l_I, 2\pi-\theta_I)$, were added for $(l_I,\theta_I)$ where $\theta_I > \pi - 5 w_\theta$. 
The entropy--scaled BADF will be used in IBI updates.

Due to the differentiability of the KDE probability distribution functions, we can compute the derivatives of the potential energy directly from the derivatives of the probability distributions. 
The derivatives of the bond--angle distributions are:

\begin{align}
    \pdv{\widehat{P}}{l} &= -\frac{1}{N}\sum\limits_{I=1}^{N} 
        \frac{K_H (\mathbf{x}-{{\mathbf{x}}_{I1}})}{\sin \theta_I} \left(\frac{l-l_{I1}}{w_l^2} \right) + \frac{K_H (\mathbf{x}-{{\mathbf{x}}_{I2}})}{\sin \theta_I} \left(\frac{l-l_{I2}}{w_l^2} \right)\label{eq:dpdl} \\ 
    \pdv{\widehat{P}}{\theta} &= -\frac{1}{N}\sum\limits_{I=1}^{N} 
        \frac{K_H (\mathbf{x} - \mathbf{x}_{I1})}{\sin \theta_I} \left(\frac{\theta - \theta_I}{w_\theta^2} \right) + 
        \frac{K_H (\mathbf{x} - \mathbf{x}_{I2})}{\sin \theta_I} \left(\frac{\theta - \theta_I}{w_\theta^2} \right)\label{eq:dpdq}\\ 
    \begin{split}
    \pdv{\widehat{P}}{\theta}{l} &= -\frac{1}{N}\sum\limits_{I=1}^{N} 
        \frac{K_H (\mathbf{x}-\mathbf{x}_{I1})}{\sin \theta_I} \left(\frac{l-l_{I1}}{w_l^2} \right) \left(\frac{\theta -\theta_I}{w_\theta^2} \right)  \\
        & \hspace{45pt} + \frac{K_H (\mathbf{x}-\mathbf{x}_{I2})}{\sin \theta_I} \left(\frac{l-l_{I2}}{w_l^2} \right) \left(\frac{\theta -\theta_I}{w_\theta^2} \right)\label{eq:dpdlq}
    \end{split}
\end{align}

Eqs.~(\ref{eq:dpdl}--\ref{eq:dpdlq}) are computationally expensive and computational cost scales with the total number of grid points multiplied by the number of sampled angle triplets.
While KDEs can be accelerated using fast Fourier transforms, it can be accomplished more straightforwardly on graphics processors. 
The code that calculated the KDE--based BADF is available on Github.\cite{ba2022}

\subsection{IBI training}
A set of 15 CG systems were generated by placing 25 chains of 80 beads, where each bead represents a C$_2$H$_4$ monomer, in a periodic simulation box via a random walk.
The simulation box size was selected so that the density $\rho$~=~0.785~g/cm$^3$, which is consistent with the reported density of amorphous PE.\cite{swan1960polyethylene,chiang1961equilibrium}
Next, the systems were relaxed using the HB potential developed by Li et al.\cite{li2019systematic}, in which a short 2--ps NVE run was performed using a displacement--limiting integrator and temperature rescaling to 600~K to eliminate nearly overlapping beads introduced by the random walk, followed by a 50--ns simulation performed in the isotropic isothermal--isobaric (NPT) ensemble at 600~K and 1~atm to equilibrate the system in the melt.
We then ramped the system temperature down to 500~K over 25~ns, which was followed by another 50--ns equilibration at 500~K in the isotropic NPT ensemble.

We subsequently reverse--mapped the systems into atomistic resolution following the method used in the previous work.\cite{agrawal2016pressure}
The atomistic interactions for PE were modeled using the polymer consistent force field (PCFF).\cite{maple1988derivation}
Then, we equilibrated the systems in the anisotropic NPT ensemble for 1~ns at 500~K and ramped the system temperature down to 300~K over 2~ns.
Finally, the systems were equilibrated at 300~K and 1~atm for 2 more nanoseconds in the anisotropic NPT ensemble. 
These 15 equilibrated atomistic systems were used as target systems.
All MD simulations in this work were performed using the Large--scale Atomic/Molecular Massively Parallel Simulator (LAMMPS). \cite{plimpton1995fast}

The average density of the equilibrated target systems is 0.758~$\pm$~0.002~g/cm$^3$ on a 95\% confidence interval.
To verify that the systems remained amorphous, we calculated their crystallinity using the local $p_2$ order parameter, the definition of which is adopted from the work by Yi et al.\cite{yi2013molecular}, written as $p_{2,i} = \left < \frac{3\cos^2 \theta_{ij} - 1}{2} \right >_j$.
The final crystallinity was 0.0019~$\pm$~0.0023.
After equilibration, the trajectories of the target atomistic systems were sampled each picosecond over a 1--ns NVT simulation performed at 300~K.
The target distribution functions were then calculated using KDE as described previously.
The grid size for the RDF was 0.01~\AA, while the grid size for the entropy--scaled BADF was 0.008~\AA~and 0.013 rad, for the bond--length and bond--angle, respectively.

To reduce the equilibration time needed in IBI iterations, we pre--equilibrated the CG systems at 300~K with HB potential to minimize the structural difference between CG systems and target atomistic systems. 
After being equilibrated at 500~K, the CG systems were then quenched to 300~K over 1~ns under isotropic NPT ensemble to minimize crystallization, which was followed by another 1--ns 300~K equilibration in the isotropic NPT ensemble.
These 15 pre--equilibrated CG systems are used in the following IBI iterations.

The potentials of mean force were used as initial guesses for the CG potential functions by computing Boltzmann inversions of the structural distribution functions. 
Using the KDE representation, we directly calculate the initial potentials and their derivatives:

\begin{align}
    U_{0}(r)                         &= -k_B T\ln g_*\\
    U_{0}(l, \theta)                 &= -k_B T\ln \widehat{P}_*\\
    \pdv{U_0}{l}         (l, \theta) &=  -\frac{k_B T}{\widehat{P}_*} \pdv{\widehat{P}_*}{l} \\
    \pdv{U_0}{\theta}    (l, \theta) &=  -\frac{k_B T}{\widehat{P}_*} \pdv{\widehat{P}_*}{\theta} \\
    \pdv{U_0}{l}{\theta} (l, \theta) &=  -\frac{k_B T}{\widehat{P}_*} \left( \pdv{\widehat{P}_*}{l}{\theta} -\frac{1}{\widehat{P}_*} \pdv{\widehat{P}_*}{l} \pdv{\widehat{P}_*}{\theta} \right)
\end{align}
where $k_B$ is the Boltzmann constant, $g_*$ and $\widehat{P}_*$ represent the RDF and entropy--scaled BADF of the target systems, respectively.
The accurate analytical energy derivatives allow bicubic interpolation to interpolate $\mathcal{V}(l, \theta)$ and its derivatives in the $l-\theta$ space.
The bicubic interpolation provides $C^1$ continuity in the energy potential.

\begin{figure}[H]
  \hspace*{-0.25in}
  \includegraphics{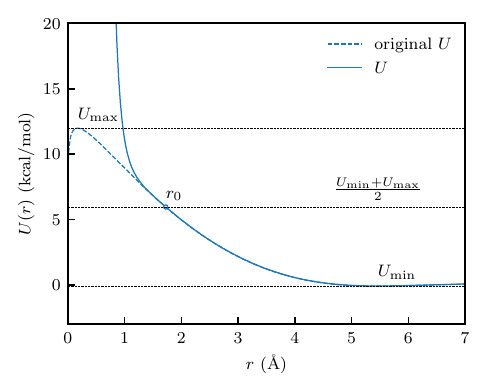}
  \caption{Demonstration of pair energy extrapolation that avoids finite energy barrier at $r=0$. }
  \label{fig:U_extrap}
\end{figure}

Two cases require correction of unphysical artifacts in the pair potentials.
The first case arises as $g(r)\rightarrow 0$ at small values of $r$ due to the excluded volume around each bead, leading to an undefined pair energy, which we replace with a short--range interaction of the form, $a r^{-9} + b r + c$.
The short--range interaction is applied for $r < r_0$, the poorly sampled region where less than a thousand pairs were observed, equivalent to $g(r) < 1e-4$ in our systems.
The parameters, $a$, $b$, and $c$ are then computed to enforce $C^2$ continuity in $U(r)$ at $r_0$.
The second case arises when the bandwidth, $w_r$, is large compared with the grid spacing $\Delta r$, which results in a spurious peak in the pair distribution function at $r = 0$ that is caused when the kernel function decays to zero more slowly than the $r_k^2$ in the kernel estimate of the RDF given in Eq.~\eqref{eq:kde_rdf}.
To resolve the issue, we used a similar extrapolation method as the first case with $r_0$ chosen so that $U(r_0) = \mathrm{max}(U)/2$, as shown in Figure~\ref{fig:U_extrap}.

Low--probability regions, where less than fifty data points were observed or $\widehat{P}(l_k,\theta_k)<10^{-5} (\mathrm{rad} \cdot \mathrm{\AA})^{-1}$, in the BADF also cause numerical difficulties when computing the potential energy and its updates.
Zero or extremely small values in the BADF lead to either undefined values or erratic energy updates that can slow or preclude the convergence of the CG potential.
Therefore, we extrapolate the bond--angle energy in the low--probability regions.
Grid points $k$, such that $\widehat{P}(l_k,\theta_k)<10^{-5} (\mathrm{rad} \cdot \mathrm{\AA})^{-1}$, are placed in a set $\mathcal T$ and all remaining grid points are placed in a set $\mathcal A$.
Next, grid point $k$ with the largest number of known neighboring points in set $\mathcal A$ is selected.
The neighbors of point $k$ are defined as all points $j$ such that $\vert l_k - l_j \vert < \delta_l$ and $\vert \theta_k - \theta_j \vert < \delta_\theta$.
The energy and derivatives at point $k$, are then recomputed using biquadratic extrapolation from the known neighboring points and point $k$ is moved to set $\mathcal A$.
The patch size of the neighboring points was selected as $\delta_l = 0.192$~\AA{} and $\delta_\theta = 0.176$~rad.
This procedure is repeated until set $\mathcal T$ is empty.

After generating initial potentials, we sampled the RDF and entropy--scaled BADF over 20~ps after a 20--ps equilibration.
We subsequently calculated $g(r)$ and $\widehat{P}(l, \theta)$ defined in Eq.~(\ref{eq:kde_rdf}--\ref{eq:badf_kernel}), which are then used to update the initial guesses of the potential.

Similar to the initial guesses, the energy updates are calculated with the target and current iteration distribution functions and their derivatives:

\begin{align}
 	\Delta U(r)                                         &= -\alpha{k_B}T  \ln \frac{g_*}{g_i}  \\
    \Delta U(l, \theta)                                 &= -\gamma{k_B}T  \ln \frac{\widehat{P}_*}{\widehat{P}_i}  \\
    \pdv{\Delta U}{l} (l, \theta)                       &= -\gamma{k_B}T \left( \pdv{\widehat{P}_*}{l}      \frac{1}{\widehat{P}_*} - \pdv{\widehat{P}_i}{l}      \frac{1}{\widehat{P}_i} \right) \\
    \pdv{\Delta U}{\theta} (l, \theta)                  &= -\gamma{k_B}T \left( \pdv{\widehat{P}_*}{\theta} \frac{1}{\widehat{P}_*} - \pdv{\widehat{P}_i}{\theta} \frac{1}{\widehat{P}_i} \right) \\
    \begin{split}
        \pdv{\Delta U}{l}{\theta}  (l, \theta)          &=  \gamma{k_B}T \Bigg[ \frac{1}{{\widehat{P}_*}^2} \left( \pdv{\widehat{P}_*}{\theta} \pdv{\widehat{P}_*}{l} - \pdv{\widehat{P}_*}{l}{\theta} \widehat{P}_* \right)  \\
        &\hspace{40pt}                                                        - \frac{1}{{\widehat{P}_i}^2} \left( \pdv{\widehat{P}_i}{\theta} \pdv{\widehat{P}_i}{l} - \pdv{\widehat{P}_i}{l}{\theta} \widehat{P}_i \right) \Bigg]
    \end{split}
\end{align}
where the coefficients $\alpha$ and $\gamma$ are scaling factors for $U(r)$ and $U(l, \theta)$, respectively, we took values of 0.15 and 0.1 for $\alpha$ and $\gamma$ to facilitate convergence.
The potential values are evaluated at the same grid points as their corresponding distribution functions.
Similar to the initial pair potentials, unphysical artifacts can appear in the updated pair potentials as well, which are corrected by the same approach.

Following the energy updates, we iteratively adjusted the updated pair potentials to ensure that the resulting pressure does not deviate far from atmospheric pressure with the method used by Wang et al.\cite{wang2009comparative}

\begin{align}
    \Delta V_{pc}(r) = A \left( 1- \frac{r}{r_c} \right)
\end{align}
where $r_c$ is the cutoff distance of non--bonded interaction, and coefficient A can be calculated as:

\begin{align}
    -\left [ \frac{2 \pi N \rho }{3 r_c} \int_0^{r_c} r^3 g(r) dr \right ] A \approx \Delta P V
\end{align}

After the pressure correction, we conduct the sampling process mentioned earlier for error evaluation.
In each iteration, after the distribution functions sampled from the updated potentials, we evaluate the representation error and the sampling error in each iteration in the same way as reported by Liu and Oswald.\cite{liu2019coarse}
The representation error of the RDF at iteration $i$ is defined as:

\begin{align}
    \epsilon_r^{(i)} = \frac{||g_i(r) - g_*(r)||_2}{||g_*(r)||_2}
\end{align}
where $||g_i(r)||$ denotes the $L_2$ norm of $g_i(r)$. 
The sampling error is defined as:

\begin{align}
    \epsilon_s^{(i)} = \frac{\sqrt{(||w_i(r)||_2)^2 + (||w_*(r)||_2)^2}}{||g_*(r)||_2}
\end{align}
where $w_i(r)$ and $w_*(r)$ are the confidence interval width of $g_i(r)$, defined as:

\begin{align}
    w(r) = t^*\frac{\sigma(r)}{\sqrt{n}}
\end{align}
where $t^*$ is the critical value for the $t$-distribution with a 95\% confidence interval and $\sigma(r)$ is the sample standard deviation, and $n$ is the number of systems at current iteration.
The error of $P(l, \theta)$ is evaluated in the same way.
We continue to the energy update of the next iteration if $\epsilon_r > \epsilon_s$, or increase the number of included CG systems if otherwise.
The trainings are considered to be converged if the maximum number of CG systems is reached.
For the trainings of the current work, we started with 5 CG systems and the maximum number was 15.

\section{Results and discussion}

\begin{figure}[tbp]
    \hspace*{-0.25in}
    \includegraphics{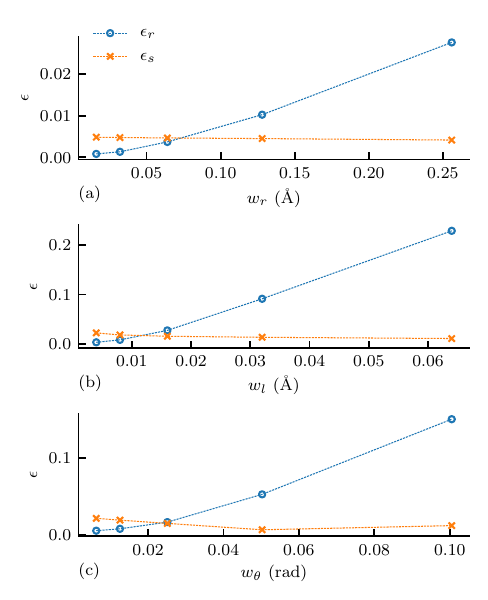}
    \caption{Relationship between sampling and representation errors with (a) pair distance bandwidth, (b) bond--length bandwidth, and (c) bond--angle bandwidth.}
    \label{fig:bw_error}
\end{figure}

To determine appropriate baseline KDE bandwidth parameters, we first compare the representation and sampling errors of the target structural distributions sampled from atomistic simulations.
The representation error measures the loss of detail incurred by increasing the KDE bandwidth and is computed as the $L_2$ norm of the difference between the distributions calculated at a particular bandwidth and the reference bandwidth.
The sampling error measures the average deviation of the distributions sampled across different systems.
The definitions of these errors for the radial distribution functions are:

\begin{align}
  & \epsilon_{r}=\frac{\left\| g_{w_r}-g_0  \right\|_2}{\left\| g_0\right\|_2}, \\
  & \epsilon_{s}=\frac{\left\| \sigma_{w_r} \right\|_2}{\left\| g_{w_r}\right\|_2},
\end{align}
where $g_{w_r}$ is the RDF calculated using bandwidth $w_r$, $g_0$ is the RDF calculated at the reference bandwidth $w_{r_0} = 0.01$~\AA, and $\sigma_{w_r}$ is the point--wise standard deviation of $g_{w_r}$ calculated from the 15 different target atomistic systems.
The representation and sampling errors of the bond--angle probability distributions are calculated as:

\begin{align}
    & \epsilon_{r} = \left\|P_{H} - P_{H_{0}}\right\|_2 \\
    & \epsilon_{s}= \int{ \sigma_{H}dl d\theta},
\end{align}
where $H$ represents the bandwidth matrix, defined by $w_l$ and $w_{\theta}$, and $\sigma_H$ denotes the standard deviation of the probability distributions computed using a given bandwidth matrix.
The bond--length and bond--angle bandwidths used for the reference bandwidth matrix are $w_{l_0} = 0.002$~\AA~ and $w_{\theta_0}= 0.0031~\mathrm{rad}$, respectively.
As shown in Figure~\ref{fig:bw_error}, increasing the kernel bandwidth parameters reduces the sampling errors but results in substantially larger increases in the representation errors.
The baseline bandwidth parameters were chosen where the representation and sampling errors have equal magnitudes: $w_r^* = 0.07$~\AA\, $w_l^* = 0.016$~\AA, and $w_{\theta}^* = 0.021~\mathrm{rad}$.

Subsequently, we conducted IBI training varying different bandwidths individually in the target distributions and studied the effect of each bandwidth on the resulting potentials. 
For $w_r$ study, we had 4 trainings using different $w_r$ with constant $w_l$ and $w_{\theta}$, e.g., $(n w^*_r, w^*_l, w^*_{\theta})$ where n = 1, 2, 4, and 8.
Studies of $w_l$ and $w_\theta$ were conducted similarly.
Therefore, there were 10 training conducted in total since $(w^*_r, w^*_l, w^*_{\theta})$ is shared in all three studies.

It is important to note that we used the baseline bandwidth set $(w^*_r, w^*_l, w^*_{\theta})$ in the CG simulations of all the trainings while changing the bandwidths of target distributions.
If the bandwidths used in training are the same as the ones in the target distributions, the smoothing of the CG and target distributions will negate each other and the CG models would be trained to reproduce the same target structure, i.e., all the converged potentials will produce almost the same distribution functions if smoothed with the same bandwidths.
Also, training with large bandwidths will have uniqueness issues; i.e., a set of strongly smoothed target distributions can be produced by more than one set of potentials due to the over--smoothed distribution functions of CG systems.

\begin{figure*}[tbp]
    \hspace*{-0.25in}
    \includegraphics{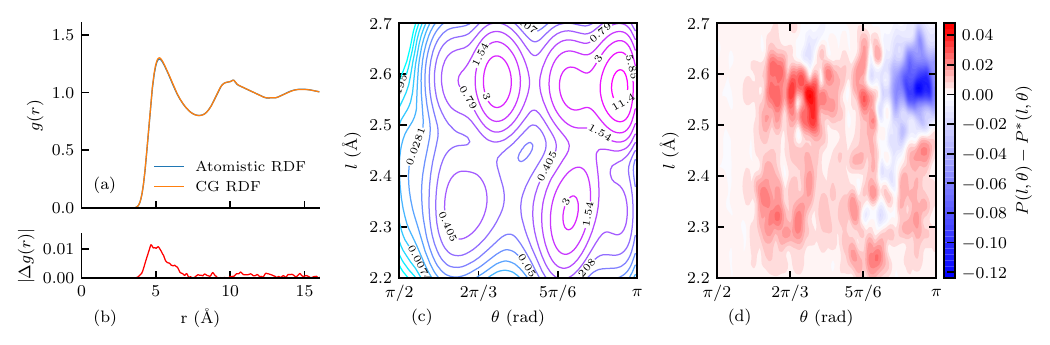}
    \caption{The training results of the bandwidth $(w^*_r, w^*_l, w^*_{\theta})$ case ($w_r = 0.07$~\AA, $w_l = 0.016 $~\AA, $w_{\theta} = 0.021$) are shown, 
        (a) shows the RDF of CG systems and atomistic target systems, 
        (b) shows the difference between the RDF of CG systems and atomistic target systems, 
        (c) shows P$(l, \theta)$ of CG systems, 
        and (d) shows the P$(l, \theta)$ difference between CG systems and atomistic target systems, the units of P$(l, \theta)$ are $1 / (\mathrm{\AA} \cdot \mathrm{rad})$.}
    \label{fig:df_compare}
\end{figure*}

\begin{figure*}[tbp]
    \hspace*{-0.25in}
    \includegraphics{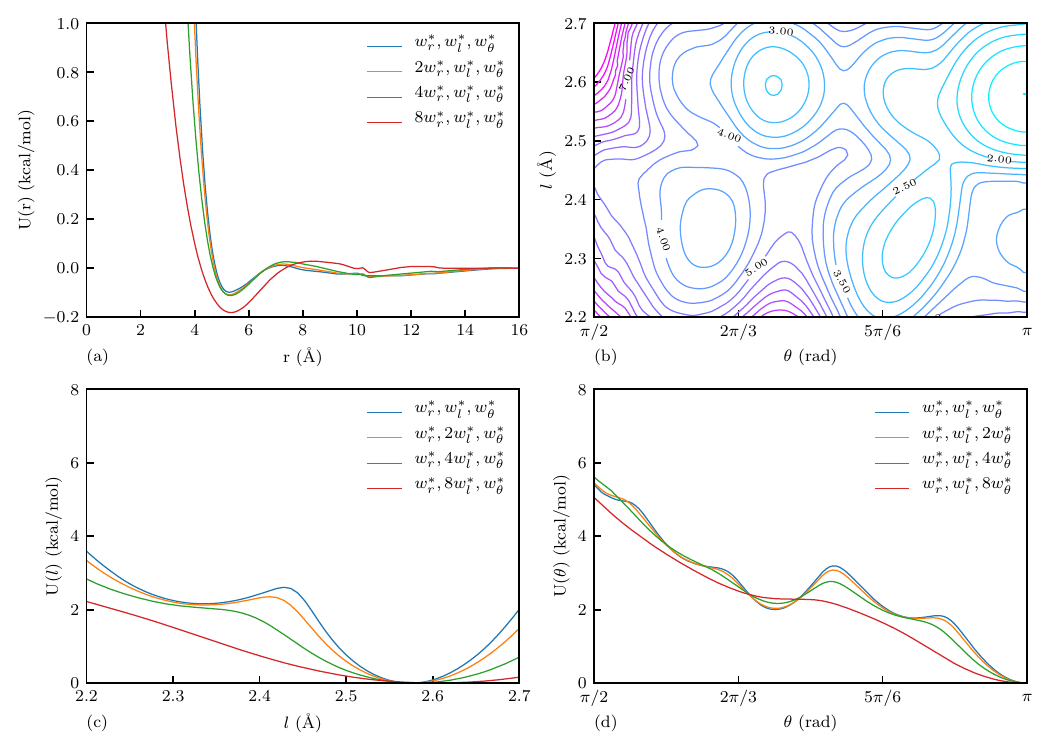}
    \caption{(a) Pair potentials trained at increasing $w_r$.
    (b) Bond--angle potential $U(l, \theta)$ trained at $(w^*_r, w^*_l, w^*_{\theta})$ bandwidth.
    (c) expectation projection of bond--angle potentials in $l$-direction, $U_l(l) = \int_0^\pi U P d\theta / \int_0^\pi P d\theta$. 
    (d) expectation projection of bond--angle potentials in $\theta$-direction, $U_{\theta}(\theta) = \int_{l_\mathrm{min}}^{l_\mathrm{max}} U P d l / \int_{l_\mathrm{min}}^{l_\mathrm{max}} P d l$.}
    \label{fig:potentials}
\end{figure*}

The trained CG models successfully reproduced the target distributions in all cases. 
A comparison between the target and CG distribution function for the $(w^*_r, w^*_l, w^*_{\theta})$ case is shown in Figure~\ref{fig:df_compare} as an example.
In addition, the pair potentials of different cases are presented in Figure~\ref{fig:potentials}(a). 
The bond--angle potential of the case $(w^*_r, w^*_l, w^*_{\theta})$ is presented in Figure~\ref{fig:potentials}(b) as an example. 
It can observed that there are four major potential wells corresponding to the 4 peaks shown in Figure~\ref{fig:df_compare}(c), representing different local conformations of PE chains.
Expectation projection of bond--angle potentials in $l$- and $\theta$-direction, as defined in the caption, are presented in Figure~\ref{fig:potentials}(c) and (d), respectively.
It is shown that as larger bandwidths are used, the BA potential would have less prominent peaks and crests, indicating that the energy barriers between the conformations are smaller.
It is not recommended to use $w_l > 4w_l^*$ or $w_\theta > 4w_\theta^*$ if accurate representation of local conformation is needed, since some gauche conformations would not be produced due to disappeared corresponding energy wells.

Following the method of Li et al.~\cite{li2019systematic}, the influence of the bandwidth parameters on the stable integration time step was determined by finding the largest time step for which the energy drift over a 1--ps NVE simulation was less than 1\% of the initial kinetic energy of the system.
Table \ref{tab:perf} compares the performance of CG models of different bandwidth cases.
It can be observed that the potentials trained in larger bandwidth cases can tolerate larger critical time steps and hence have better computational performance.
It is verified that the highest--frequency mode is the bond vibration, since the critical time step significantly increases when the $l$-direction stiffness is reduced by using target distributions with larger $w_l$.
From the perspective of time step--per--second, our CG--BA model has a performance of 29.78 time step/s with an 8000--monomer system and one MPI task, which is about 91\% of the performance of the HB model.
The slowdown is due to more computationally expensive bicubic interpolation compared to the simple linear interpolation used in the HB model and the cost induced by extra MPI gathering and communication of bond information.

\begin{table}[h!]
  \centering
  \begin{tabular}{cccccc} \toprule
      $w_r$ & $w_l$ & $w_\theta$ & $\Delta t_c$ & CPU & Speed \\ 
      (\AA{}) & (\AA) & (rad)  &  (fs) & (s) & up\\ \midrule
       0.07 & 0.016 & 0.021      &  4 & 1.05  & $\times$47 \\
       0.14 & 0.016 & 0.021      &  6 & 0.70  & $\times$70 \\
       0.28 & 0.016 & 0.021      &  4 & 1.05  & $\times$47 \\
       0.56 & 0.016 & 0.021      &  6 & 0.70  & $\times$71 \\ 
       0.07 & 0.032 & 0.021      &  4 & 1.05  & $\times$47 \\
       0.07 & 0.064 & 0.021      &  7 & 0.60  & $\times$83 \\
       0.07 & 0.128 & 0.021      & 15 & 0.28  & $\times$177 \\
       0.07 & 0.016 & 0.042      &  4 & 1.05  & $\times$47 \\
       0.07 & 0.016 & 0.084      &  6 & 0.70  & $\times$71 \\
       0.07 & 0.016 & 0.168      &  6 & 0.70  & $\times$71 \\
      \multicolumn{3}{c}{HB}     &  5 & 0.76  & $\times$65 \\  
      \multicolumn{3}{c}{all-atom (PCFF)} & 1 & 49.5 & -- \\\bottomrule
  \end{tabular}
  \caption{Performance comparison of the models calibrated using different 
           bandwidth parameters, where $\Delta t_c$ is the critical time step, 
           CPU is the wall time required per monomer per nanosecond for an 
           NPT simulation performed on a single core on an Intel (R) Xeon (R) E5-2680 v4, 
           and the speed-up factor is the relative performance to an all--atom simulation.}
    \label{tab:perf}
\end{table}

\begin{figure*}[tbp]
    \hspace*{-0.25in}
    \includegraphics{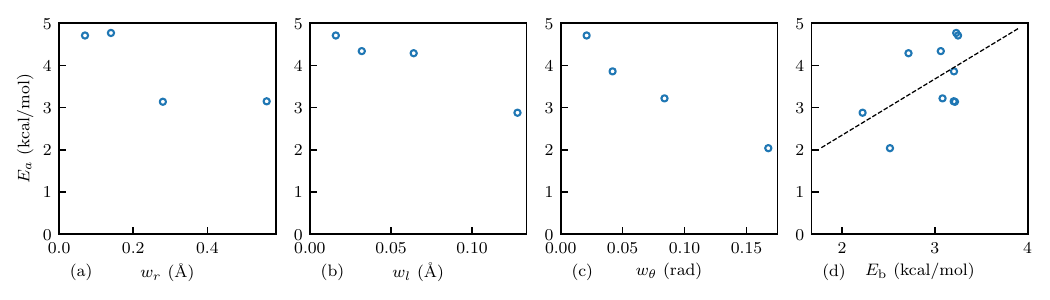}
    \caption{Figure (a), (b), and (c) show how activation energy of crystalline chain diffusion changes with respect to $w_r$, $w_l$, and $w_\theta$, respectively.
    Figure (d) shows the correlation between the activation energy and the energy barrier between the states as demonstrated by the dashed line.}
    \label{fig:Ea_Emin_bw}
\end{figure*}

To understand the effect of bandwidths on the crystalline phase chain diffusion behavior of semicrystalline polymers, we calculated the activation energy of the potentials trained with different targets.
We constructed a crystal system in a periodic simulation cell with x, y, and z dimensions of 5.8, 5.3, and 20.5~nm, respectively. 
A total of 154 chains were placed parallel to the z-axis and are bonded across the periodic boundaries, representing an infinite lamellar thickness. 
The system were simulated in the anisotropic NPT ensemble for 50~ns at 1~atm, and various temperatures, namely 240, 270, 300, 330, and 370~K. 
Mean--squared displacements were measured to quantify diffusion.
The diffusion coefficients were calculated with $D = \frac{g(t)}{2t} |_{t=50 \mathrm{ns}}$ and the activation energy $E_a$ was calculated with the Arrhenius relationship.
The results are shown in Figure~\ref{fig:Ea_Emin_bw}.

Figure~\ref{fig:Ea_Emin_bw}(a), (b), and (c) show the relationship between the activation energy and the $w_r$, $w_l$, and $w_{\theta}$, respectively.
While increasing $w_r$ also tends to reduce the activation energy of crystalline diffusion, $w_l$ and $w_\theta$ demonstrate a stronger correlation with $E_a$, i.e., potentials from larger bandwidth trainings have a reduced activation energy of crystalline diffusion.
We attribute the lower activation energy to the reduced energy barriers in $U(l, \theta)$ caused by stronger smoothing from larger bandwidths.
It was shown that the defect movements are the agents of the chain--direction slip (or chain diffusion) in the crystal phases and account for the $\alpha_c$-relaxation.\cite{mowry2002atomistic}
This indicates that the state transitions of torsion angles in local chain segments, which form conformational defects and enable defect movements, would also facilitate diffusion.
The torsion angle transitions in the atomistic representation are implicitly represented by the state transitions of bond--angle triplets in our CG mapping as shown in Figure~\ref{fig:bdf}, meaning that energy barrier between the local conformation states in $U(l, \theta)$ strongly affects the activation energy of crystalline diffusion.

We then identified the energy barrier $E_b$ of the BA potentials.
More specifically, for a given BA potential, we first found all the local minima (states) in $U(l, \theta)$ and identified the trans state $T$.
Then, we identified the minimum energy pathway from $T$ to all other states with the algorithm developed by Fu et al.\cite{fu2020finding} and calculated the maximum energy on the paths, which were regarded as the energy barriers of the paths.
Eventually, the average energy barriers of all the paths are used as the energy barrier $E_b$ of the BA potential.

The relationship between the activation energy $E_a$ and the energy barrier $E_b$ is shown in Figure~\ref{fig:Ea_Emin_bw}(d), where the data points follow a straight line. 
It verifies our hypothesis that the energy barrier in $U(l, \theta)$ correlates strongly with the activation energy $E_a$; and the reduction in the activation energy observed in crystalline diffusion $E_a$ is accounted by the decrease in $E_b$ due to stronger smoothing.
Because the energy landscape is smoothened by the coarse--graining, the activation energy is lowered when compared with the experimental values of 5 to 22 kcal/mol\cite{mowry2002atomistic}, and 23 to 30 kcal/mol\cite{ashcraft1976dielectric}, which will facilitate changes in local conformations.
However, the controllability on $E_b$ provided by the bandwidth parameters offers a useful tool to adjust $E_a$ of CG models for different applications.

To understand how the bandwidth parameters can be used to accelerate crystallization kinetics to allow studies to be conducted at larger length and time scales, we simulated the crystallization process with 5 randomly generated systems using the trained CG potentials.
For each system, 280 chains with 285 CG beads representing a molecular weight of 7980 g/mol were randomly placed in a cubic simulation box with side lengths of 10.6~nm.
The chain length was selected so that polymer chains can crystallize into large crystallites and chains can pass through a crystallite multiple times and form loops.
We then ran simulations in the anisotropic NPT ensemble at 300~K, 1~atm for 100~ns, during which the systems crystallized.

\begin{figure}[tbp]
    \hspace*{-0.25in}
    \includegraphics{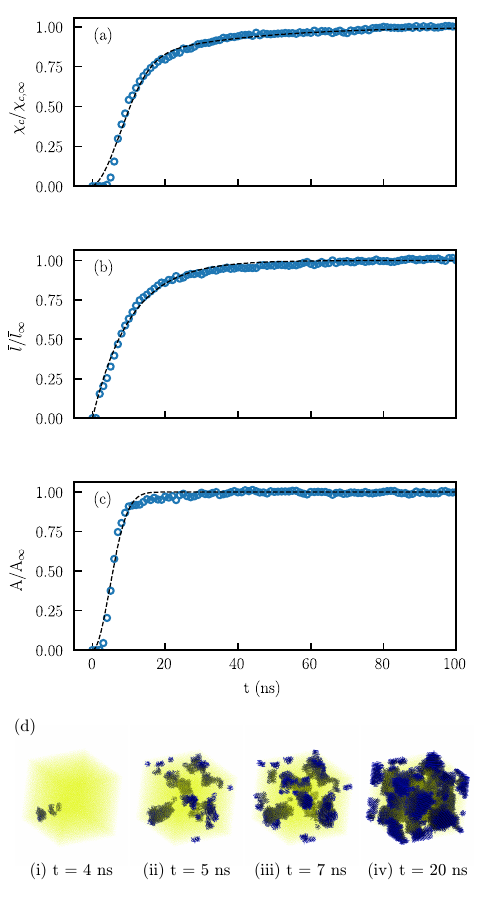}
    \caption{Evolution of (a) relative crystalline phase volume,
                          (b) relative mean crystal stem length, 
                          (c) relative crystallite cross--sectional area perpendicular to the stem direction, and 
                          (d) visualization of crystallite growth at selected time steps.
            The dashed line in plots (a) is fit to Velisaris--Seferis\cite{velisaris1986crystallization}, whereas the ones in plot (b) and (c) are fit to Avrami model\cite{avrami1939kinetics}.}
    \label{fig:D_l_vs_time}
\end{figure}

The crystallization of semicrystalline polymer consists of two processes.
In primary crystallization, which is commonly modeled with the Avrami equation\cite{avrami1939kinetics}, crystallites grow radially from nuclei and form spherulites.
This process majorly consists of the growth of lamellae in the direction perpendicular to the stems (transverse direction).
In secondary crystallization, lamellae thicken by reorganizing the polymer chains in adjacent amorphous phases in a manner similar to chain slips.
We identified the characteristics of crystallites with the following procedures.
After dividing the simulation boxes into bins of 7~\AA{}, we calculated the $p_2$ parameter of the beads with a cutoff radius of 14~\AA{}, and then identified the crystallites by finding the contiguous clusters of bins with average $p_2$ parameter greater than 0.4.
To measure the secondary crystallization rate, we then calculated the average stem length in crystallites.
To quantify the crystallite growth in the transverse direction, cross--sectional area of crystallites in the transverse plane were also estimated.
The results are reported in Figure~\ref{fig:D_l_vs_time}.

The crystallite growth follows a typical polymer crystallization process.
The transverse growth starts and saturates quickly, providing the initial growth of crystallites, combined with stem growth that is responsible for the second--stage crystallite growth.
Such a trend can also be observed from the rendered images in Figure~\ref{fig:D_l_vs_time}(d).
In image (i) and (ii), the system evolved from two nuclei to multiple small crystallites, meaning multiple nucleation sites formed which indicates sufficiently large domain size.
The primary crystallization dominates the process from image (ii) to image (iii) which illustrates the system state only 2~ns later.
Image (iv) demonstrates a fully crystallized state and after which no significant change was observed.
The half--time of crystallization, defined as the time to reach 50\% of relative crystallinity, is about 10~ns; 
when comparing to the experimental results by Zhuravlev et al.\cite{zhuravlev2016crystallization}, we can roughly estimate that the acceleration is at most 100 times.
More accurate estimation is impossible due to missing data for isothermal crytallization of polyethylene at 300~K.
The CG model with accelerated crystallization kinetics is by no means ideal to reproduce realistic time scale, but does provide a viable and efficient approach to study the crystallization processes.

\begin{figure}[tbp]
    \hspace*{-0.25in}
    \includegraphics{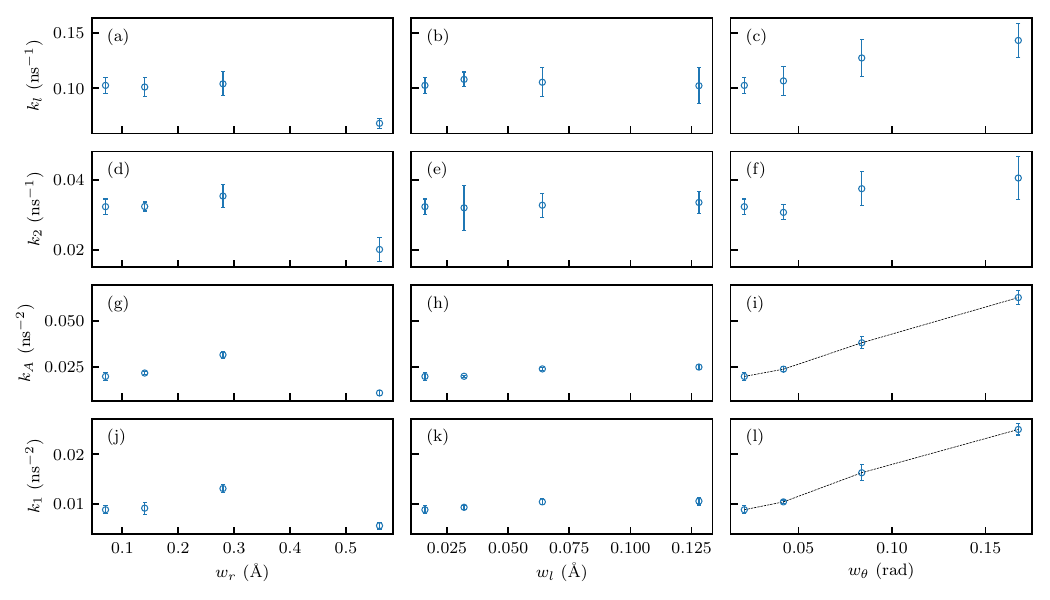}
    \caption{Sensitivity of bandwidth parameters $w_r$, $w_l$, and $w_{\theta}$ on crystal growth rate parameters $k_l$, $k_A$, $k_1$, and $k_2$.
             The error bars indicate the 90\% confidence intervals.
             The strongest trends are highlighted using dashed lines.}
    \label{fig:growth_fit_params}
\end{figure}

Multiple models are available to describe crystallite growth.
The Avrami equation\cite{avrami1939kinetics} is commonly used to model crystallization kinetics, defined as:
\begin{equation}
    \chi_c/\chi_{c, \infty} = 1 - \exp(-k t^n_1)
\end{equation}
where $\chi_c$ and $\chi_{c, \infty}$ denote the crystallinity and final crystallinity, $k$ is the crystallization rate, and $n$ is the Avrami exponent.
The Avrami model assumes isotropic growth hence can not account for the secondary crystallization process.
Among the works\cite{hillier1965modified,ravindranath1993polymer,tobin1974theory,tobin1976theory,tobin1977theory,malkin1984general,urbanovici1990new,urbanovici1996isothermal} that attempt to take the secondary crystallization into consideration, the Hay model\cite{chen2016effect,phillipson2016effect} and Velisaris--Seferis model\cite{velisaris1986crystallization} were shown to generate the best fit according to Kelly and Jenkins.\cite{kelly2022modeling} 
We therefore selected the Velisaris--Seferis model because it converges to a finite value at infinite time.
It is defined as:
\begin{equation}
    \chi_c / \chi_{c, \infty} = w_1(1 - e^{-k_1 t^n_1}) + w_2(1- e^{-k_2 t^n_2})
\end{equation}
where $k_1$ and $k_2$ signify the primary and secondary crystallization rates, respectively, $\chi_c$ denotes the crystallinity, $\chi_{c, \infty}$ is the final crystallinity, $w_1$ and $w_2$ are the weights of primary and secondary crystallization and sum to one, $n_1$ and $n_2$ are the Avrami exponents for primary and secondary crystallization.
We selected $n_1 = 2$ and $n_2=1$ since the primary crystallization and the secondary crystallization are volume nucleation with two--dimensional and one--dimensional growth, respectively.
We fitted the data in Figure~\ref{fig:D_l_vs_time}(a) with the Velisaris--Seferis model to analyze the crystal growth.
Naturally, the data shown in Figure~\ref{fig:D_l_vs_time}(b) and (c) are fitted with the Avrami model.
To distinguish the fitted parameters, the transverse growth rate is denoted by $k_A$ and the stem growth rate is denoted by $k_l$, which has the same physical meaning as $k_1$ and $k_2$, respectively.
All the fittings matched with the data points well as shown in Figure~\ref{fig:D_l_vs_time} and the parameters are reported in Figure~\ref{fig:growth_fit_params}.

It can be observed that the trends in the row of $k_l$ and $k_A$ agree well with the row of $k_2$ and $k_1$ (e.g., the trend in Figure~\ref{fig:growth_fit_params}(a) is similar to the one in Figure~\ref{fig:growth_fit_params}(d)), respectively, which validates the fittings and post--processing since the parameters with the same physical implications are affected similarly.
When considering the time to reach a certain relative crystsallinity level, the primary crystallization is generally 2 to 5 times faster than the secondary crystallization, consistent with the conclusion by Banks et al.\cite{banks1963crystallization} that secondary crystallization is less than one order of magnitude slower than the primary crystallization.
The parameter $w_r$ reduces the secondary crystallization rate when $w_r$ is large, due to the deeper pair potential well in the ($8w_r$, $w_l$, $w_{\theta}$) case posing higher energy barrier to chain slips.
$w_l$ has little effect on the crystallization kinetics in the investigated range.
Increasing $w_\theta$ greatly increases $k_A$ and $k_1$; the increase of $k_A$ and $k_1$ is attributed to a weaker constraint at $\theta = \pi$ resulting from larger $w_\theta$, which facilitates the folding of the chains and accelerates the crystallite growth in the transverse direction.

\section{Conclusions}
In this work, we developed an extension to the IBI method that considers correlated multi--variable intramolecular interactions.
This method also incorporates the KDE representation in distributions and energy formulations, which provides controllability to the smoothness of the resulting potential functions and the characteristics of the resulting models.

We trained 10 CG models of PE with different kernel bandwidths using this method.
The models successfully reproduced the correlated distributions of bond--length and bond--angle of PE with an $L_2$ norm error of about 1\%.
The conformation states of bond--angle triplets, which represents the underlying atomistic conformations, were reproduced with appropriate bandwidths.
We found that the activation energy of crystalline chain diffusion of the CG model is determined by the energy barrier $E_\mathrm{b}$ between the conformational states. 
By adjusting the bandwidths of the target distributions, we could tune the magnitude of the peaks and crests in the $l-\theta$ energy landscape, giving us controllability towards the system dynamics.
Increasing the bandwidths will accelerate the dynamics by smoothing out the peaks and crests at the cost of weakened ability to accurately reproduce the local conformations; it will also reduce the stiffness of the BA potential and the maximum frequency of the model, leading to increased critical time step and computational performance.
By adjusting the bandwidth, one could focus on accurately reproducing the local conformations, or efficiently cover long time scale in simulations

In the study of isothermal crystallization of supercooled PE, the models successfully crystallized the systems and reproduced the primary and secondary crystallization kinetics.
The model trained with larger $w_r$ would slow down the secondary crystallization rates because deep pair potential well creates resistance for chain slip, whereas $w_l$ have little effect on the crystallization kinetics within the investigated range.
Additionally, increasing $w_\theta$ greatly increases the primary crystallization rate, indicating the flexibility of chains has a significant impact on the crystallite growth in the transverse direction.
The CG model presented in this work shows controllable characteristics in diffusion dynamics and crystallization kinetics.
We look forward to further testing the capability of the model in simulating mechanical properties.

\section{Acknowledgements}
The authors acknowledge support from the National Science Foundation (NSF) under award Division of Civil, Mechanical, and Manufacturing Innovation (CMMI) 1653830.
The authors acknowledge Research Computing at Arizona State University for providing HPC resources that have contributed to the research results reported within this paper.

\bibliography{references}

\providecommand{\latin}[1]{#1}
\makeatletter
\providecommand{\doi}
  {\begingroup\let\do\@makeother\dospecials
  \catcode`\{=1 \catcode`\}=2 \doi@aux}
\providecommand{\doi@aux}[1]{\endgroup\texttt{#1}}
\makeatother
\providecommand*\mcitethebibliography{\thebibliography}
\csname @ifundefined\endcsname{endmcitethebibliography}
  {\let\endmcitethebibliography\endthebibliography}{}
\begin{mcitethebibliography}{57}
\providecommand*\natexlab[1]{#1}
\providecommand*\mciteSetBstSublistMode[1]{}
\providecommand*\mciteSetBstMaxWidthForm[2]{}
\providecommand*\mciteBstWouldAddEndPuncttrue
  {\def\EndOfBibitem{\unskip.}}
\providecommand*\mciteBstWouldAddEndPunctfalse
  {\let\EndOfBibitem\relax}
\providecommand*\mciteSetBstMidEndSepPunct[3]{}
\providecommand*\mciteSetBstSublistLabelBeginEnd[3]{}
\providecommand*\EndOfBibitem{}
\mciteSetBstSublistMode{f}
\mciteSetBstMaxWidthForm{subitem}{(\alph{mcitesubitemcount})}
\mciteSetBstSublistLabelBeginEnd
  {\mcitemaxwidthsubitemform\space}
  {\relax}
  {\relax}

\bibitem[Nielsen \latin{et~al.}(2004)Nielsen, Lopez, Srinivas, and
  Klein]{nielsen2004coarse}
Nielsen,~S.~O.; Lopez,~C.~F.; Srinivas,~G.; Klein,~M.~L. Coarse grain models
  and the computer simulation of soft materials. \emph{J. Phys. Condens.
  Matter} \textbf{2004}, \emph{16}, R481\relax
\mciteBstWouldAddEndPuncttrue
\mciteSetBstMidEndSepPunct{\mcitedefaultmidpunct}
{\mcitedefaultendpunct}{\mcitedefaultseppunct}\relax
\EndOfBibitem
\bibitem[Peter and Kremer(2009)Peter, and Kremer]{peter2009multiscale}
Peter,~C.; Kremer,~K. Multiscale simulation of soft matter systems--from the
  atomistic to the coarse-grained level and back. \emph{Soft Matter}
  \textbf{2009}, \emph{5}, 4357--4366\relax
\mciteBstWouldAddEndPuncttrue
\mciteSetBstMidEndSepPunct{\mcitedefaultmidpunct}
{\mcitedefaultendpunct}{\mcitedefaultseppunct}\relax
\EndOfBibitem
\bibitem[Liu and Oswald(2019)Liu, and Oswald]{liu2019coarse}
Liu,~M.; Oswald,~J. Coarse--grained molecular modeling of the microphase
  structure of polyurea elastomer. \emph{Polymer} \textbf{2019}, \emph{176},
  1--10\relax
\mciteBstWouldAddEndPuncttrue
\mciteSetBstMidEndSepPunct{\mcitedefaultmidpunct}
{\mcitedefaultendpunct}{\mcitedefaultseppunct}\relax
\EndOfBibitem
\bibitem[Liu \latin{et~al.}(2023)Liu, Ye, and Oswald]{liu2023coarse}
Liu,~M.; Ye,~J.; Oswald,~J. Coarse-grained molecular simulation of the role of
  curing rates on the structure and strength of polyurea. \emph{Comput. Mater.
  Sci.} \textbf{2023}, \emph{230}, 112428\relax
\mciteBstWouldAddEndPuncttrue
\mciteSetBstMidEndSepPunct{\mcitedefaultmidpunct}
{\mcitedefaultendpunct}{\mcitedefaultseppunct}\relax
\EndOfBibitem
\bibitem[Salerno \latin{et~al.}(2016)Salerno, Agrawal, Perahia, and
  Grest]{salerno2016resolving}
Salerno,~K.~M.; Agrawal,~A.; Perahia,~D.; Grest,~G.~S. Resolving dynamic
  properties of polymers through coarse-grained computational studies.
  \emph{Phys. Rev. Lett.} \textbf{2016}, \emph{116}, 058302\relax
\mciteBstWouldAddEndPuncttrue
\mciteSetBstMidEndSepPunct{\mcitedefaultmidpunct}
{\mcitedefaultendpunct}{\mcitedefaultseppunct}\relax
\EndOfBibitem
\bibitem[Hoy and Karayiannis(2013)Hoy, and Karayiannis]{hoy2013simple}
Hoy,~R.~S.; Karayiannis,~N.~C. Simple model for chain packing and
  crystallization of soft colloidal polymers. \emph{Phys. Rev. E}
  \textbf{2013}, \emph{88}, 012601\relax
\mciteBstWouldAddEndPuncttrue
\mciteSetBstMidEndSepPunct{\mcitedefaultmidpunct}
{\mcitedefaultendpunct}{\mcitedefaultseppunct}\relax
\EndOfBibitem
\bibitem[Nguyen \latin{et~al.}(2015)Nguyen, Smith, Hoy, and
  Karayiannis]{nguyen2015effect}
Nguyen,~H.~T.; Smith,~T.~B.; Hoy,~R.~S.; Karayiannis,~N.~C. Effect of chain
  stiffness on the competition between crystallization and glass-formation in
  model unentangled polymers. \emph{J. Chem. Phys.} \textbf{2015}, \emph{143},
  144901\relax
\mciteBstWouldAddEndPuncttrue
\mciteSetBstMidEndSepPunct{\mcitedefaultmidpunct}
{\mcitedefaultendpunct}{\mcitedefaultseppunct}\relax
\EndOfBibitem
\bibitem[Paul \latin{et~al.}(1995)Paul, Yoon, and Smith]{paul1995optimized}
Paul,~W.; Yoon,~D.~Y.; Smith,~G.~D. An optimized united atom model for
  simulations of polymethylene melts. \emph{J. Chem. Phys.} \textbf{1995},
  \emph{103}, 1702--1709\relax
\mciteBstWouldAddEndPuncttrue
\mciteSetBstMidEndSepPunct{\mcitedefaultmidpunct}
{\mcitedefaultendpunct}{\mcitedefaultseppunct}\relax
\EndOfBibitem
\bibitem[Waheed \latin{et~al.}(2002)Waheed, Lavine, and
  Rutledge]{waheed2002molecular}
Waheed,~N.; Lavine,~M.; Rutledge,~G. Molecular simulation of crystal growth in
  n-eicosane. \emph{J. Chem. Phys.} \textbf{2002}, \emph{116}, 2301--2309\relax
\mciteBstWouldAddEndPuncttrue
\mciteSetBstMidEndSepPunct{\mcitedefaultmidpunct}
{\mcitedefaultendpunct}{\mcitedefaultseppunct}\relax
\EndOfBibitem
\bibitem[Waheed \latin{et~al.}(2005)Waheed, Ko, and
  Rutledge]{waheed2005molecular}
Waheed,~N.; Ko,~M.; Rutledge,~G. Molecular simulation of crystal growth in long
  alkanes. \emph{Polymer} \textbf{2005}, \emph{46}, 8689--8702\relax
\mciteBstWouldAddEndPuncttrue
\mciteSetBstMidEndSepPunct{\mcitedefaultmidpunct}
{\mcitedefaultendpunct}{\mcitedefaultseppunct}\relax
\EndOfBibitem
\bibitem[Yi \latin{et~al.}(2013)Yi, Locker, and Rutledge]{yi2013molecular}
Yi,~P.; Locker,~C.~R.; Rutledge,~G.~C. Molecular dynamics simulation of
  homogeneous crystal nucleation in polyethylene. \emph{Macromolecules}
  \textbf{2013}, \emph{46}, 4723--4733\relax
\mciteBstWouldAddEndPuncttrue
\mciteSetBstMidEndSepPunct{\mcitedefaultmidpunct}
{\mcitedefaultendpunct}{\mcitedefaultseppunct}\relax
\EndOfBibitem
\bibitem[Zubova \latin{et~al.}(2017)Zubova, Strelnikov, Balabaev, Savin, Mazo,
  and Manevich]{zubova2017coarse}
Zubova,~E.; Strelnikov,~I.; Balabaev,~N.; Savin,~A.; Mazo,~M.; Manevich,~L.
  Coarse-grained polyethylene: 1. The simplest model for the orthorhombic
  crystal. \emph{Polym. Sci. Ser. A} \textbf{2017}, \emph{59}, 149--158\relax
\mciteBstWouldAddEndPuncttrue
\mciteSetBstMidEndSepPunct{\mcitedefaultmidpunct}
{\mcitedefaultendpunct}{\mcitedefaultseppunct}\relax
\EndOfBibitem
\bibitem[Strelnikov \latin{et~al.}(2017)Strelnikov, Zubova, Mazo, and
  Manevich]{strelnikov2017coarse}
Strelnikov,~I.; Zubova,~E.; Mazo,~M.; Manevich,~L. Coarse-grained polyethylene:
  Including cross terms in bonded interactions and introducing anisotropy into
  the model for the orthorhombic crystal. \emph{Polym. Sci. Ser. A}
  \textbf{2017}, \emph{59}, 242--252\relax
\mciteBstWouldAddEndPuncttrue
\mciteSetBstMidEndSepPunct{\mcitedefaultmidpunct}
{\mcitedefaultendpunct}{\mcitedefaultseppunct}\relax
\EndOfBibitem
\bibitem[Lavine \latin{et~al.}(2003)Lavine, Waheed, and Rutledge]{Lavine2003}
Lavine,~M.~S.; Waheed,~N.; Rutledge,~G.~C. {Molecular dynamics simulation of
  orientation and crystallization of polyethylene during uniaxial extension}.
  \emph{Polymer} \textbf{2003}, \emph{44}, 1771--1779\relax
\mciteBstWouldAddEndPuncttrue
\mciteSetBstMidEndSepPunct{\mcitedefaultmidpunct}
{\mcitedefaultendpunct}{\mcitedefaultseppunct}\relax
\EndOfBibitem
\bibitem[Yamamoto(2004)]{Yamamoto2004}
Yamamoto,~T. {Molecular dynamics modeling of polymer crystallization from the
  melt}. \emph{Polymer} \textbf{2004}, \emph{45}, 1357--1364\relax
\mciteBstWouldAddEndPuncttrue
\mciteSetBstMidEndSepPunct{\mcitedefaultmidpunct}
{\mcitedefaultendpunct}{\mcitedefaultseppunct}\relax
\EndOfBibitem
\bibitem[Ko \latin{et~al.}(2004)Ko, Waheed, Lavine, and Rutledge]{Ko2004}
Ko,~M.~J.; Waheed,~N.; Lavine,~M.~S.; Rutledge,~G.~C. {Characterization of
  polyethylene crystallization from an oriented melt by molecular dynamics
  simulation}. \emph{Journal of Chemical Physics} \textbf{2004}, \emph{121},
  2823--2832\relax
\mciteBstWouldAddEndPuncttrue
\mciteSetBstMidEndSepPunct{\mcitedefaultmidpunct}
{\mcitedefaultendpunct}{\mcitedefaultseppunct}\relax
\EndOfBibitem
\bibitem[Yamamoto(2013)]{Yamamoto2013}
Yamamoto,~T. {Molecular dynamics of polymer crystallization revisited:
  Crystallization from the melt and the glass in longer polyethylene}.
  \emph{Journal of Chemical Physics} \textbf{2013}, \emph{139}, 54903\relax
\mciteBstWouldAddEndPuncttrue
\mciteSetBstMidEndSepPunct{\mcitedefaultmidpunct}
{\mcitedefaultendpunct}{\mcitedefaultseppunct}\relax
\EndOfBibitem
\bibitem[Paajanen \latin{et~al.}(2019)Paajanen, Vaari, and
  Verho]{paajanen2019crystallization}
Paajanen,~A.; Vaari,~J.; Verho,~T. Crystallization of cross-linked polyethylene
  by molecular dynamics simulation. \emph{Polymer} \textbf{2019}, \emph{171},
  80--86\relax
\mciteBstWouldAddEndPuncttrue
\mciteSetBstMidEndSepPunct{\mcitedefaultmidpunct}
{\mcitedefaultendpunct}{\mcitedefaultseppunct}\relax
\EndOfBibitem
\bibitem[Sliozberg \latin{et~al.}(2018)Sliozberg, Yeh, Kr\"{o}ger, Masser,
  Lenhart, and Andzelm]{sliozberg2018ordering}
Sliozberg,~Y.~R.; Yeh,~I.-C.; Kr\"{o}ger,~M.; Masser,~K.~A.; Lenhart,~J.~L.;
  Andzelm,~J.~W. Ordering and crystallization of entangled polyethylene melts
  under uniaxial tension: a molecular dynamics study. \emph{Macromolecules}
  \textbf{2018}, \emph{51}, 9635--9648\relax
\mciteBstWouldAddEndPuncttrue
\mciteSetBstMidEndSepPunct{\mcitedefaultmidpunct}
{\mcitedefaultendpunct}{\mcitedefaultseppunct}\relax
\EndOfBibitem
\bibitem[Larini \latin{et~al.}(2010)Larini, Lu, and Voth]{larini2010multiscale}
Larini,~L.; Lu,~L.; Voth,~G.~A. The multiscale coarse-graining method. VI.
  Implementation of three-body coarse-grained potentials. \emph{J. Chem. Phys.}
  \textbf{2010}, \emph{132}, 164107\relax
\mciteBstWouldAddEndPuncttrue
\mciteSetBstMidEndSepPunct{\mcitedefaultmidpunct}
{\mcitedefaultendpunct}{\mcitedefaultseppunct}\relax
\EndOfBibitem
\bibitem[Poursina \latin{et~al.}(2011)Poursina, Bhalerao, Flores, Anderson, and
  Laederach]{poursina2011strategies}
Poursina,~M.; Bhalerao,~K.~D.; Flores,~S.~C.; Anderson,~K.~S.; Laederach,~A.
  Strategies for articulated multibody-based adaptive coarse grain simulation
  of RNA. \emph{methods Enzymol.} \textbf{2011}, \emph{487}, 73--98\relax
\mciteBstWouldAddEndPuncttrue
\mciteSetBstMidEndSepPunct{\mcitedefaultmidpunct}
{\mcitedefaultendpunct}{\mcitedefaultseppunct}\relax
\EndOfBibitem
\bibitem[Munson and Singh(1997)Munson, and Singh]{munson1997statistical}
Munson,~P.~J.; Singh,~R.~K. Statistical significance of hierarchical multi-body
  potentials based on Delaunay tessellation and their application in
  sequence-structure alignment. \emph{Protein Sci.} \textbf{1997}, \emph{6},
  1467--1481\relax
\mciteBstWouldAddEndPuncttrue
\mciteSetBstMidEndSepPunct{\mcitedefaultmidpunct}
{\mcitedefaultendpunct}{\mcitedefaultseppunct}\relax
\EndOfBibitem
\bibitem[Li and Liang(2005)Li, and Liang]{li2005geometric}
Li,~X.; Liang,~J. Geometric cooperativity and anticooperativity of three-body
  interactions in native proteins. \emph{Proteins: Struct., Funct., Bioinf.}
  \textbf{2005}, \emph{60}, 46--65\relax
\mciteBstWouldAddEndPuncttrue
\mciteSetBstMidEndSepPunct{\mcitedefaultmidpunct}
{\mcitedefaultendpunct}{\mcitedefaultseppunct}\relax
\EndOfBibitem
\bibitem[Ejtehadi \latin{et~al.}(2004)Ejtehadi, Avall, and
  Plotkin]{ejtehadi2004three}
Ejtehadi,~M.; Avall,~S.; Plotkin,~S. Three-body interactions improve the
  prediction of rate and mechanism in protein folding models. \emph{Proc. Natl.
  Acad. Sci.} \textbf{2004}, \emph{101}, 15088--15093\relax
\mciteBstWouldAddEndPuncttrue
\mciteSetBstMidEndSepPunct{\mcitedefaultmidpunct}
{\mcitedefaultendpunct}{\mcitedefaultseppunct}\relax
\EndOfBibitem
\bibitem[Krishna \latin{et~al.}(2010)Krishna, Ayton, and Voth]{krishna2010role}
Krishna,~V.; Ayton,~G.~S.; Voth,~G.~A. Role of protein interactions in defining
  HIV-1 viral capsid shape and stability: a coarse-grained analysis.
  \emph{Biophys. J.} \textbf{2010}, \emph{98}, 18--26\relax
\mciteBstWouldAddEndPuncttrue
\mciteSetBstMidEndSepPunct{\mcitedefaultmidpunct}
{\mcitedefaultendpunct}{\mcitedefaultseppunct}\relax
\EndOfBibitem
\bibitem[Schapotschnikow and Vlugt(2009)Schapotschnikow, and
  Vlugt]{schapotschnikow2009understanding}
Schapotschnikow,~P.; Vlugt,~T.~J. Understanding interactions between capped
  nanocrystals: Three-body and chain packing effects. \emph{J. Chem. Phys.}
  \textbf{2009}, \emph{131}, 124705\relax
\mciteBstWouldAddEndPuncttrue
\mciteSetBstMidEndSepPunct{\mcitedefaultmidpunct}
{\mcitedefaultendpunct}{\mcitedefaultseppunct}\relax
\EndOfBibitem
\bibitem[Fukunaga \latin{et~al.}(2002)Fukunaga, Takimoto, and
  Doi]{fukunaga2002coarse}
Fukunaga,~H.; Takimoto,~J.-i.; Doi,~M. A coarse-graining procedure for flexible
  polymer chains with bonded and nonbonded interactions. \emph{J. Chem. Phys.}
  \textbf{2002}, \emph{116}, 8183--8190\relax
\mciteBstWouldAddEndPuncttrue
\mciteSetBstMidEndSepPunct{\mcitedefaultmidpunct}
{\mcitedefaultendpunct}{\mcitedefaultseppunct}\relax
\EndOfBibitem
\bibitem[Reith \latin{et~al.}(2003)Reith, P{\"u}tz, and
  M{\"u}ller-Plathe]{reith2003deriving}
Reith,~D.; P{\"u}tz,~M.; M{\"u}ller-Plathe,~F. Deriving effective mesoscale
  potentials from atomistic simulations. \emph{J. Comput. Chem.} \textbf{2003},
  \emph{24}, 1624--1636\relax
\mciteBstWouldAddEndPuncttrue
\mciteSetBstMidEndSepPunct{\mcitedefaultmidpunct}
{\mcitedefaultendpunct}{\mcitedefaultseppunct}\relax
\EndOfBibitem
\bibitem[McCabe \latin{et~al.}(2014)McCabe, Korb, and Cole]{mccabe2014kernel}
McCabe,~P.; Korb,~O.; Cole,~J. Kernel Density Estimation Applied to Bond
  Length, Bond Angle, and Torsion Angle Distributions. \emph{J. Chem. Inf.
  Model.} \textbf{2014}, \emph{54}, 1284--1288\relax
\mciteBstWouldAddEndPuncttrue
\mciteSetBstMidEndSepPunct{\mcitedefaultmidpunct}
{\mcitedefaultendpunct}{\mcitedefaultseppunct}\relax
\EndOfBibitem
\bibitem[Li \latin{et~al.}(2019)Li, Agrawal, and Oswald]{li2019systematic}
Li,~Y.; Agrawal,~V.; Oswald,~J. Systematic coarse-graining of semicrystalline
  polyethylene. \emph{J. Polym. Sci. Part B Polym. Phys.} \textbf{2019},
  \emph{57}, 331--342\relax
\mciteBstWouldAddEndPuncttrue
\mciteSetBstMidEndSepPunct{\mcitedefaultmidpunct}
{\mcitedefaultendpunct}{\mcitedefaultseppunct}\relax
\EndOfBibitem
\bibitem[Ye and Oswald(2022)Ye, and Oswald]{CGBA2022}
Ye,~J.; Oswald,~J. CGBA-potentials. 2022;
  \url{https://doi.org/10.5281/zenodo.6590569}\relax
\mciteBstWouldAddEndPuncttrue
\mciteSetBstMidEndSepPunct{\mcitedefaultmidpunct}
{\mcitedefaultendpunct}{\mcitedefaultseppunct}\relax
\EndOfBibitem
\bibitem[Ye and Oswald(2022)Ye, and Oswald]{ba2022}
Ye,~J.; Oswald,~J. ba-probability. 2022;
  \url{https://doi.org/10.5281/zenodo.6590571}\relax
\mciteBstWouldAddEndPuncttrue
\mciteSetBstMidEndSepPunct{\mcitedefaultmidpunct}
{\mcitedefaultendpunct}{\mcitedefaultseppunct}\relax
\EndOfBibitem
\bibitem[Swan(1960)]{swan1960polyethylene}
Swan,~P.~R. Polyethylene specific volume, crystallinity, and glass transition.
  \emph{J. Polym. Sci.} \textbf{1960}, \emph{42}, 525--534\relax
\mciteBstWouldAddEndPuncttrue
\mciteSetBstMidEndSepPunct{\mcitedefaultmidpunct}
{\mcitedefaultendpunct}{\mcitedefaultseppunct}\relax
\EndOfBibitem
\bibitem[Chiang and Flory(1961)Chiang, and Flory]{chiang1961equilibrium}
Chiang,~R.; Flory,~P. Equilibrium between Crystalline and Amorphous Phases in
  Polyethylene1. \emph{J. Am. Chem. Soc.} \textbf{1961}, \emph{83},
  2857--2862\relax
\mciteBstWouldAddEndPuncttrue
\mciteSetBstMidEndSepPunct{\mcitedefaultmidpunct}
{\mcitedefaultendpunct}{\mcitedefaultseppunct}\relax
\EndOfBibitem
\bibitem[Agrawal \latin{et~al.}(2016)Agrawal, Peralta, Li, and
  Oswald]{agrawal2016pressure}
Agrawal,~V.; Peralta,~P.; Li,~Y.; Oswald,~J. A pressure-transferable
  coarse-grained potential for modeling the shock Hugoniot of polyethylene.
  \emph{J. Chem. Phys.} \textbf{2016}, \emph{145}, 104903\relax
\mciteBstWouldAddEndPuncttrue
\mciteSetBstMidEndSepPunct{\mcitedefaultmidpunct}
{\mcitedefaultendpunct}{\mcitedefaultseppunct}\relax
\EndOfBibitem
\bibitem[Maple \latin{et~al.}(1988)Maple, Dinur, and
  Hagler]{maple1988derivation}
Maple,~J.~R.; Dinur,~U.; Hagler,~A.~T. Derivation of force fields for molecular
  mechanics and dynamics from ab initio energy surfaces. \emph{Proc. Natl.
  Acad. Sci. U. S. A.} \textbf{1988}, \emph{85}, 5350--5354\relax
\mciteBstWouldAddEndPuncttrue
\mciteSetBstMidEndSepPunct{\mcitedefaultmidpunct}
{\mcitedefaultendpunct}{\mcitedefaultseppunct}\relax
\EndOfBibitem
\bibitem[Plimpton(1995)]{plimpton1995fast}
Plimpton,~S. Fast parallel algorithms for short-range molecular dynamics.
  \emph{J. Comput. Phys.} \textbf{1995}, \emph{117}, 1--19\relax
\mciteBstWouldAddEndPuncttrue
\mciteSetBstMidEndSepPunct{\mcitedefaultmidpunct}
{\mcitedefaultendpunct}{\mcitedefaultseppunct}\relax
\EndOfBibitem
\bibitem[Wang \latin{et~al.}(2009)Wang, Junghans, and
  Kremer]{wang2009comparative}
Wang,~H.; Junghans,~C.; Kremer,~K. Comparative atomistic and coarse-grained
  study of water: What do we lose by coarse-graining? \emph{Eur. Phys. J. E}
  \textbf{2009}, \emph{28}, 221--229\relax
\mciteBstWouldAddEndPuncttrue
\mciteSetBstMidEndSepPunct{\mcitedefaultmidpunct}
{\mcitedefaultendpunct}{\mcitedefaultseppunct}\relax
\EndOfBibitem
\bibitem[Mowry and Rutledge(2002)Mowry, and Rutledge]{mowry2002atomistic}
Mowry,~S.~W.; Rutledge,~G.~C. Atomistic simulation of the $\alpha_c$-Relaxation
  in crystalline polyethylene. \emph{Macromolecules} \textbf{2002}, \emph{35},
  4539--4549\relax
\mciteBstWouldAddEndPuncttrue
\mciteSetBstMidEndSepPunct{\mcitedefaultmidpunct}
{\mcitedefaultendpunct}{\mcitedefaultseppunct}\relax
\EndOfBibitem
\bibitem[Fu \latin{et~al.}(2020)Fu, Chen, Wang, Chai, Shao, Cai, and
  Chipot]{fu2020finding}
Fu,~H.; Chen,~H.; Wang,~X.; Chai,~H.; Shao,~X.; Cai,~W.; Chipot,~C. Finding an
  optimal pathway on a multidimensional free-energy landscape. \emph{J. Chem.
  Inf. Model.} \textbf{2020}, \emph{60}, 5366--5374\relax
\mciteBstWouldAddEndPuncttrue
\mciteSetBstMidEndSepPunct{\mcitedefaultmidpunct}
{\mcitedefaultendpunct}{\mcitedefaultseppunct}\relax
\EndOfBibitem
\bibitem[Ashcraft and Boyd(1976)Ashcraft, and Boyd]{ashcraft1976dielectric}
Ashcraft,~C.~R.; Boyd,~R.~H. A dielectric study of molecular relaxation in
  oxidized and chlorinated polyethylenes. \emph{J. Polym. Science, Polym. Phys.
  Ed.} \textbf{1976}, \emph{14}, 2153--2193\relax
\mciteBstWouldAddEndPuncttrue
\mciteSetBstMidEndSepPunct{\mcitedefaultmidpunct}
{\mcitedefaultendpunct}{\mcitedefaultseppunct}\relax
\EndOfBibitem
\bibitem[Avrami(1939)]{avrami1939kinetics}
Avrami,~M. Kinetics of phase change. I General theory. \emph{J. Chem. Phys.}
  \textbf{1939}, \emph{7}, 1103--1112\relax
\mciteBstWouldAddEndPuncttrue
\mciteSetBstMidEndSepPunct{\mcitedefaultmidpunct}
{\mcitedefaultendpunct}{\mcitedefaultseppunct}\relax
\EndOfBibitem
\bibitem[Velisaris and Seferis(1986)Velisaris, and
  Seferis]{velisaris1986crystallization}
Velisaris,~C.~N.; Seferis,~J.~C. Crystallization kinetics of
  polyetheretherketone (PEEK) matrices. \emph{Polym. Eng. Sci.} \textbf{1986},
  \emph{26}, 1574--1581\relax
\mciteBstWouldAddEndPuncttrue
\mciteSetBstMidEndSepPunct{\mcitedefaultmidpunct}
{\mcitedefaultendpunct}{\mcitedefaultseppunct}\relax
\EndOfBibitem
\bibitem[Zhuravlev \latin{et~al.}(2016)Zhuravlev, Madhavi, Lustiger, Androsch,
  and Schick]{zhuravlev2016crystallization}
Zhuravlev,~E.; Madhavi,~V.; Lustiger,~A.; Androsch,~R.; Schick,~C.
  Crystallization of polyethylene at large undercooling. \emph{ACS Macro Lett.}
  \textbf{2016}, \emph{5}, 365--370\relax
\mciteBstWouldAddEndPuncttrue
\mciteSetBstMidEndSepPunct{\mcitedefaultmidpunct}
{\mcitedefaultendpunct}{\mcitedefaultseppunct}\relax
\EndOfBibitem
\bibitem[Hillier(1965)]{hillier1965modified}
Hillier,~I. Modified avrami equation for the bulk crystallization kinetics of
  spherulitic polymers. \emph{J. Polym. Sci. Part A Polym. Phys.}
  \textbf{1965}, \emph{3}, 3067--3078\relax
\mciteBstWouldAddEndPuncttrue
\mciteSetBstMidEndSepPunct{\mcitedefaultmidpunct}
{\mcitedefaultendpunct}{\mcitedefaultseppunct}\relax
\EndOfBibitem
\bibitem[Ravindranath and Jog(1993)Ravindranath, and
  Jog]{ravindranath1993polymer}
Ravindranath,~K.; Jog,~J. Polymer crystallization kinetics: Poly (ethylene
  terephthalate) and poly (phenylene sulfide). \emph{J. Appl. Polym. Sci.}
  \textbf{1993}, \emph{49}, 1395--1403\relax
\mciteBstWouldAddEndPuncttrue
\mciteSetBstMidEndSepPunct{\mcitedefaultmidpunct}
{\mcitedefaultendpunct}{\mcitedefaultseppunct}\relax
\EndOfBibitem
\bibitem[Tobin(1974)]{tobin1974theory}
Tobin,~M.~C. Theory of phase transition kinetics with growth site impingement.
  I. Homogeneous nucleation. \emph{J. Polym. Science, Polym. Phys. Ed.}
  \textbf{1974}, \emph{12}, 399--406\relax
\mciteBstWouldAddEndPuncttrue
\mciteSetBstMidEndSepPunct{\mcitedefaultmidpunct}
{\mcitedefaultendpunct}{\mcitedefaultseppunct}\relax
\EndOfBibitem
\bibitem[Tobin(1976)]{tobin1976theory}
Tobin,~M.~C. The theory of phase transition kinetics with growth site
  impingement. II. Heterogeneous nucleation. \emph{J. Polym. Science, Polym.
  Phys. Ed.} \textbf{1976}, \emph{14}, 2253--2257\relax
\mciteBstWouldAddEndPuncttrue
\mciteSetBstMidEndSepPunct{\mcitedefaultmidpunct}
{\mcitedefaultendpunct}{\mcitedefaultseppunct}\relax
\EndOfBibitem
\bibitem[Tobin(1977)]{tobin1977theory}
Tobin,~M.~C. Theory of phase transition kinetics with growth site impingement.
  III. Mixed heterogeneous--homogeneous nucleation and nonintegral exponents of
  the time. \emph{J. Polym. Science, Polym. Phys. Ed.} \textbf{1977},
  \emph{15}, 2269--2270\relax
\mciteBstWouldAddEndPuncttrue
\mciteSetBstMidEndSepPunct{\mcitedefaultmidpunct}
{\mcitedefaultendpunct}{\mcitedefaultseppunct}\relax
\EndOfBibitem
\bibitem[Malkin \latin{et~al.}(1984)Malkin, Beghishev, Keapin, and
  Bolgov]{malkin1984general}
Malkin,~A.~Y.; Beghishev,~V.; Keapin,~I.~A.; Bolgov,~S. General treatment of
  polymer crystallization kinetics—part 1. A new macrokinetic equation and
  its experimental verification. \emph{Polymer Engineering \& Science}
  \textbf{1984}, \emph{24}, 1396--1401\relax
\mciteBstWouldAddEndPuncttrue
\mciteSetBstMidEndSepPunct{\mcitedefaultmidpunct}
{\mcitedefaultendpunct}{\mcitedefaultseppunct}\relax
\EndOfBibitem
\bibitem[Urbanovici and Segal(1990)Urbanovici, and Segal]{urbanovici1990new}
Urbanovici,~E.; Segal,~E. New formal relationships to describe the kinetics of
  crystallization. \emph{Thermochimica acta} \textbf{1990}, \emph{171},
  87--94\relax
\mciteBstWouldAddEndPuncttrue
\mciteSetBstMidEndSepPunct{\mcitedefaultmidpunct}
{\mcitedefaultendpunct}{\mcitedefaultseppunct}\relax
\EndOfBibitem
\bibitem[Urbanovici \latin{et~al.}(1996)Urbanovici, Schneider, Brizzolara, and
  Cantow]{urbanovici1996isothermal}
Urbanovici,~E.; Schneider,~H.; Brizzolara,~D.; Cantow,~H. Isothermal melt
  crystallization kinetics of poly (L-lactic acid). \emph{J. Therm. Anal.
  Calorim.} \textbf{1996}, \emph{47}, 931--939\relax
\mciteBstWouldAddEndPuncttrue
\mciteSetBstMidEndSepPunct{\mcitedefaultmidpunct}
{\mcitedefaultendpunct}{\mcitedefaultseppunct}\relax
\EndOfBibitem
\bibitem[Chen \latin{et~al.}(2016)Chen, Hay, and Jenkins]{chen2016effect}
Chen,~Z.; Hay,~J.~N.; Jenkins,~M.~J. The effect of secondary crystallization on
  crystallization kinetics--Polyethylene terephthalate revisited.
  \emph{European Polymer Journal} \textbf{2016}, \emph{81}, 216--223\relax
\mciteBstWouldAddEndPuncttrue
\mciteSetBstMidEndSepPunct{\mcitedefaultmidpunct}
{\mcitedefaultendpunct}{\mcitedefaultseppunct}\relax
\EndOfBibitem
\bibitem[Phillipson \latin{et~al.}(2016)Phillipson, Jenkins, and
  Hay]{phillipson2016effect}
Phillipson,~K.; Jenkins,~M.~J.; Hay,~J.~N. The effect of a secondary process on
  crystallization kinetics--Poly ($\epsilon$-caprolactone) revisited.
  \emph{European Polymer Journal} \textbf{2016}, \emph{84}, 708--714\relax
\mciteBstWouldAddEndPuncttrue
\mciteSetBstMidEndSepPunct{\mcitedefaultmidpunct}
{\mcitedefaultendpunct}{\mcitedefaultseppunct}\relax
\EndOfBibitem
\bibitem[Kelly and Jenkins(2022)Kelly, and Jenkins]{kelly2022modeling}
Kelly,~C.~A.; Jenkins,~M.~J. Modeling the crystallization kinetics of polymers
  displaying high levels of secondary crystallization. \emph{Polym. J.}
  \textbf{2022}, \emph{54}, 249--257\relax
\mciteBstWouldAddEndPuncttrue
\mciteSetBstMidEndSepPunct{\mcitedefaultmidpunct}
{\mcitedefaultendpunct}{\mcitedefaultseppunct}\relax
\EndOfBibitem
\bibitem[Banks \latin{et~al.}(1963)Banks, Gordon, Roe, and
  Sharples]{banks1963crystallization}
Banks,~W.; Gordon,~M.; Roe,~R.; Sharples,~A. The crystallization of
  polyethylene I. \emph{Polymer} \textbf{1963}, \emph{4}, 61--74\relax
\mciteBstWouldAddEndPuncttrue
\mciteSetBstMidEndSepPunct{\mcitedefaultmidpunct}
{\mcitedefaultendpunct}{\mcitedefaultseppunct}\relax
\EndOfBibitem
\end{mcitethebibliography}
\end{document}